\documentclass{eptcs}

\usepackage{bsymb,b2latex} 

\usepackage{latexsym,amssymb,amsmath,amsthm}
\usepackage{epsf}
\usepackage{graphicx}
\usepackage{subfigure}

\usepackage[usenames,dvipsnames]{color}  

\usepackage{listings}

\def\mymath#1{\relax\ifmmode#1\else$#1$\fi}

\newcommand{\theoremstart}[2]{{\bf #1 #2}}

\newcounter{defcounter}[section]
\renewcommand{\thedefcounter}{\thesection.\arabic{defcounter}}

{\refstepcounter{defcounter} \begin{trivlist} \item[]%
\theoremstart{Algorithm}{\thedefcounter}}%
{\end{trivlist}}

   {\begin{list}{--}{
   \setlength{\itemsep}{0 pt}
   \setlength{\parsep}{0 pt}
   \setlength{\topsep} {0 pt} }}
   {\end{list}}
   {\begin{list}{}{
   \setlength{\itemsep}{0 pt}
   \setlength{\parsep}{0 pt}
   \setlength{\topsep} {2 pt} }}
   {\end{list}}

\newcommand{\ignore}[1]{} 
\newcommand{\forExperts}[1]{} 
\newcommand{\comment}[1]{}

\newcommand{\prob}{{\sc ProB}}

\newdimen\mytab    \mytab=1em

\begin{document}
\theoremstyle{definition}
\newtheorem{example}{Example}

\title{Who watches the watchers:\\Validating the ProB Validation Tool\thanks{
This research is being carried out as part of the EU funded FP7 research project 287563 (Advance).}}

\def\titlerunning{Validating ProB}
\def\authorrunning{Bendisposto, Krings, Leuschel}

\author{
Jens Bendisposto \qquad\qquad Sebastian Krings \qquad\qquad Michael Leuschel
\institute{
 Institut f\"{u}r Informatik, Heinrich-Heine Universit\"{a}t D\"{u}sseldorf\\
  Universit\"{a}tsstr. 1, D-40225 D\"{u}sseldorf\\}
  \email{\{bendisposto, krings, leuschel\}@cs.uni-duesseldorf.de}
}

\maketitle

\pagestyle{empty}

\begin{abstract}
Over the years, \prob\ has moved from a tool that {\em complemented\/} proving, 
 to a development environment that is now sometimes used {\em instead of\/} proving for applications, such as exhaustive model checking or data validation. This has led to much more stringent requirements on the integrity of \prob.
In this paper we present a summary of our validation efforts for \prob, in particular within the context of the norm EN 50128 and safety critical applications in the railway domain.
\end{abstract}
\section{Introduction}
Over the years, \prob\ has moved from being a tool that {\em complemented\/} proving, 
 to being a tool that is now sometimes used {\em instead of\/} proving for certain applications. 
 \prob\  can be used as a plug-in for proof-centric development environmants like Rodin or Atelier-B, but it can also be used as a standalone development environment that focuses on validation or verification using animation and model checking.
For example, \prob\ is sometimes used to exhaustively model check B specifications which are not easily amenable to proving (e.g., where it is not
 easy to find inductive invariants).
In 2009 \prob\ started to be used for data validation within Siemens \cite{DBLP:journals/fac/LeuschelFFP11},
 replacing custom proof procedures.
In this application, complicated properties had to be validated on concrete data values for safety critical railway applications.
As this turned out to be highly successful, \prob\ is now being used for data validation by a variety of other companies (Alstom, ClearSy, Systerel) and
 for a variety of industrial end users (e.g., RATP).

All this means that we can now longer rely on the integrity of the B provers in case a bug in \prob\ prevents the detection of an error.
As such, validation of the \prob\ tool became very important and even mandatory.
Indeed, according to the railway norm EN 50128 \cite{EN50128} [Sect. 3.1], \prob\ is now being used as a tool of class T2, which mandates
 a certain amount of documentation and validation.
Our long term goal is to enable \prob\ to also be used as a tool of class T3, i.e., moving from a tool of class T2 that 
 {\em ``supports the test or verification of the design or executable code, where errors in the tool can fail to reveal defects but cannot directly create errors in the executable software''} \cite{EN50128} to a tool that
 {\em ``generates outputs which can directly or indirectly contribute to the executable code (including data) of the safety related system''} \cite{EN50128}.
 In addition to the validation efforts described in this paper, T3 would require a secondary toolchain as described in ``Checking Computations of Formal Method Tools - A Secondary Toolchain for \prob''.\footnote{To appear in proceedings of F-IDE 2014.}
In this paper we outline the various validation efforts that have been conducted in that light, and present a summary of our experiences.


\section{Architecture}

\subsection{Source Code}

The kernel of \prob\ is developed in Prolog, but many other programming languages come into play.
More precisely, the source code of \prob\
 contains more than 45,000 lines of Prolog, more than 15,000 lines of Tcl/Tk,
 more than 5,000 lines of C (for LTL and symmetry reduction),
 1,951 lines of grammar and more than 10,000 lines of Java that are expanded by the SableCC compiler generator \cite{gagnon:sablecc} into about 91,000 lines of Java for our parsers.
In addition, there are slightly more than 5,000 lines of Haskell code for the CSP parser and about 70,000 lines of
  Java code for the Rodin \cite{rodinplatform} plugin. These are the statistics for version 1.3.7-beta9 of \prob.
  
We use Git for our source code repositories, managing a history that traces back to 2005.
The repository has been ported to different SCMs multiple times, starting from an initial CVS repository, to SVN and to GIT. In addition to our main Git repositories we manage several SVN repositories for examples, test cases and documentation. These have not been moved to Git for practical reasons.
   
We have two versions of the \prob\ binary:
\begin{itemize}
 \item the ProB Tcl/Tk version, which has a GUI and
 \item the probcli binary.
  It does not depend on Tcl/Tk and can be used from the command-line, from within Makefiles and via a socket interface
   which listens to requests. The latter is used for integrating \prob\ into the Rodin Eclipse platform.
\end{itemize}

Both binaries are used for industrial applications such as data validation; the Tcl/Tk version has the additional benefit of
 providing a predicate debugger and graphical visualization of formulas.

\subsection{Important Components}
\begin{figure}
  \begin{center}

    \includegraphics[width= 11.5cm]{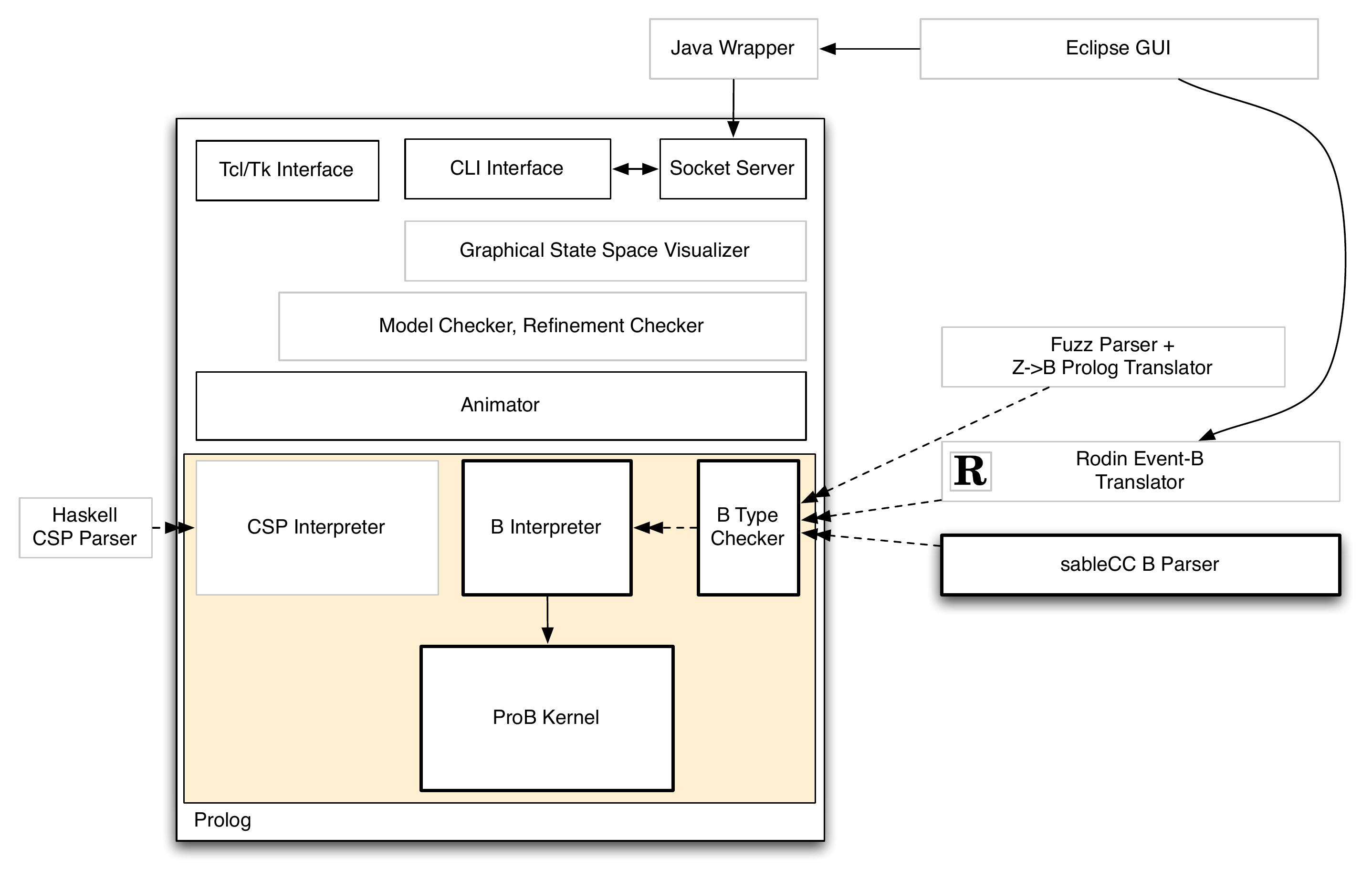}
  \end{center}
   \caption{Architecture of ProB with highlighting of critical components (for data validation)} 
\label{ProB_Architecture_STS}
\end{figure}

 A graphical representation of the \prob\ architecture can be found in Figure~\ref{ProB_Architecture_STS}.
 The most important components are the following:
 \begin{itemize}
 \item
 The \prob\ kernel deals with B's datatypes (integers, booleans, pairs, sets, ...) and the various
  operations on them (arithmetic, set operations, ...).
 \item
 The B interpreter interprets the structure of a B machine to compute the initial states, the enabled
  operations, the effect of performing an operation, etc.
 The B interpreter calls the \prob\ kernel (via the kernel\_mappings interface file).
 \item The B Type Checker reads in the parsed abstract syntax tree (AST), type checks it and
  performs certain pre-processing.
  \item the B Parser reads in a series of B files and produces an abstract syntax tree (AST) in Prolog form
   (written into a file ending with ``.prob'').
\end{itemize}
 
 Any error in those components is likely to result in a potential error during data validation or model checking, and as
  such these components need to be tested very thoroughly.

\subsubsection{ProB kernel}

The kernel supports the underlying data structures and operators of the B language:
   arithmetic, sets, relations, functions and sequences. 
As such, the specification for the ProB kernel is the reference book by 
 Abrial \cite{Abrial:BBook}, in particular Chapters 2 and 3.
As additional reference, the Atelier-B handbook \cite{atelierb40} has been used, for those operators not covered in \cite{Abrial:BBook}.
 
 Many of the laws from \cite{Abrial:BBook} are tested below, in various forms, e.g., instances are used as unit tests
  or more systematically, in Section~\ref{it:model-check} we ensure that ProB can find no counter-example to
  those laws.
 For example, on page 124 we can find the law {\lstinline[mathescape]+ union(SS) = {x| $\exists$ y . (y $\in$ SS $\wedge$ x$\in$y)}+}, which
  is added as one of the laws in the machine checking set laws.
The particular example from that page --- {\verb+u = {{0,5,2,4}, {2,4,5}, {2,1,7,5}}+} with
  {\verb+union(u) = {0,1,2,4,5,7}+} ---
    is added as a regression test to the machine checking explicit computations of laws.

Several mistakes in the B-Book were found in this process; e.g., the first definition of override on page 80 uses $p$ instead
 of $r$.
In the second definition, it should be (a,b) instead of (x,y).
On page 102, the property involving $ran(p;q) = ran(p)$ is false, it should be $ran(q)$.
  
\subsubsection{B Interpreter}

The B interpreter evaluates B predicates, expressions%
\footnote{See page 31 of \cite{Abrial:BBook} for the difference.}
 and substitutions.
As such, again the book by Abrial \cite{Abrial:BBook} is the main specification.

In particular, Chapter 1.2 is used for predicates,
 and Chapter 4 is used for substitutions.
As additional reference, the Atelier-B handbook \cite{atelierb40} has been used, for those aspects not covered in \cite{Abrial:BBook}.

For substitutions, the definitions in \cite{Abrial:BBook}
 are in form of rules for the weakest pre-condition.
\prob\ uses an interpreter which, given a predecessor state, computes all possible successor states.
(An alternate view is that \prob\ finds solutions for the before-after predicate of a substitution.)
As such, the weakest pre-condition definitions do not apply directly, and \prob\ is validated
in other ways.

\subsubsection{Parser, Type Checker and Syntax Tree Manipulations}

For the parser, the syntax of Atelier-B  \cite{atelierb40} is the reference specification.
The type checking basics are specified in \cite{Abrial:BBook}.
For the visibility rules, the tables in \cite{atelierb40} have been used as reference.


\section{Building and Testing through Continuous Integration}
\label{sec:jenkins} \label{testing}

\begin{figure}
\begin{center}
  \includegraphics[width=9cm]{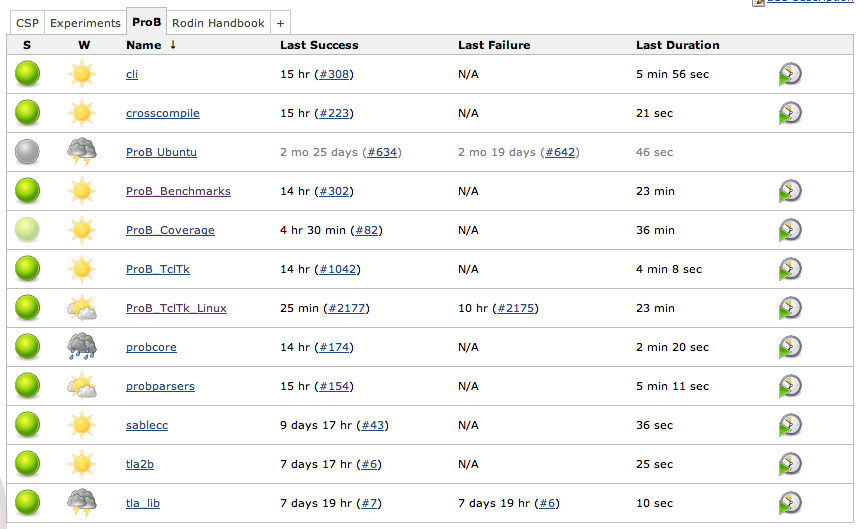} 
  
  \includegraphics[width=3.2cm]{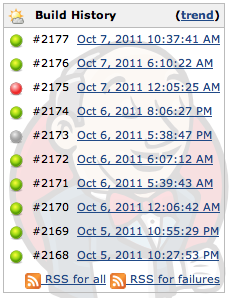} 
  \includegraphics[width=5.8cm]{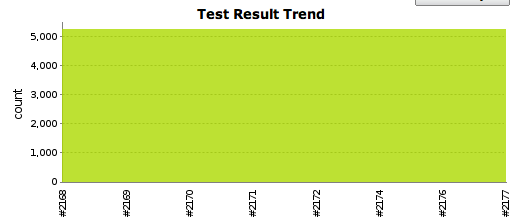} 
  \caption{Different web pages of our Jenkins server}\label{fig:jenkins}
\end{center}
\end{figure}

\prob\ contains unit tests, integration and regression tests as well as
 self-model check tests for mathematical laws (see below).
All of these tests are run automatically on our continuous integration platform ``Jenkins''.%
\footnote{See {\tt http://en.wikipedia.org/wiki/Jenkins\_(software)}.}
The tests are run on all platforms for which we ship \prob.
When a test fails, an email is sent automatically to the \prob\ development team.
The status of the tests can be monitored on the web pages of our Jenkins instance (see Figure~\ref{fig:jenkins}):
 
  {\tt http://cobra.cs.uni-duesseldorf.de/jenkins/}.

Over the years many tests have been developed.
They can be classified into unit tests for the \prob\ kernel, parser tests, type checker tests, and system tests exercising the whole \prob\ tooling infrastructure.

\subsection{Unit Tests for Kernel}
  \label{it:unit}%
  \prob\ contains over a 1,000 manually entered unit tests at the Prolog level. 
 Most of them check the
   proper functioning of the various \prob\ kernel predicates operating on B's data structures.
   For example, the Prolog implementation of the union operator is exercised to check that among others
    $\{1\} \cup \{2\}$ evaluates to $\{1,2\}$.
    
There are two situation in which we manually add tests.
The first is to add unit tests following the test-driven development process.
The second is to add tests aiming to increase code coverage as we will explain in Section \ref{sec:coverage}.

  In addition, we have developed an automatic {\em unit test generator},
    which tests the \prob\ kernel predicates for many 
    different scenarios and different set representations. 
  For example, starting from the initial Prolog call
     {\tt\small union([int(1)],[int(2)],[int(1),int(2)])},
      the test generator will derive 1358 individual unit tests.
   The various tests will use different set representations (AVL-tree or symbolic representations),
     will swap the order of the first two arguments as union is declared commutative, and will check various orderings in which
     the sets are instantiated (e.g., it could be that first the result of the union
     is known, then the second argument).
   The latter point is particularly important for the \prob\ kernel, which relies on co-routines.
   Indeed,
     we have to check that the kernel predicates behave correctly no matter in which order (partial) information is propagated.

\subsection{Validation of the parser} \label{parser_validation}

The \prob\ parser is written in Java and has been developed using SableCC.
We use about 1800 JUnit based tests to check various parts of the parser for errors.
While most of them verify the correct behavior on small examples or single expressions, some employ full scale B machines.
The unit tests are run independently on both Windows and Linux.
Several of these tests are automatically adopted to different newline characters and encodings. 

\begin{figure} 
\begin{center}
  \includegraphics[width=10.5cm]{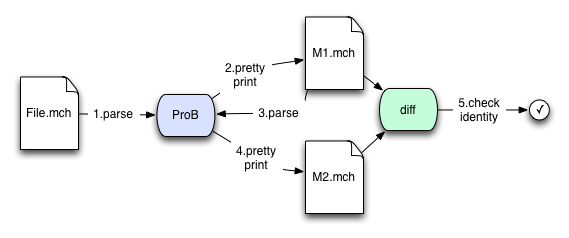} 
  \caption{Overview of the Parser Validation Process}\label{ParserValidation}
\end{center}
\end{figure}

  Additionally, we execute our parser on a large number of our regression test machines and pretty print the
   internal representation.
  We then parse the internal representation and pretty print it again,
   verifying (with {\tt diff}) that we get exactly the same result
   (see Figure~\ref{ParserValidation}).
  This type of validation can easily be applied to a large number of B machines,
   and will detect if the parser omits, reorders or modifies expressions, provided that
    the pretty printer does not compensate errors of the parser.
  On the downside, the validation will only detect those errors in machines generated by the pretty
   printer, which may prevent us from catching errors which only appear in non-pretty printed machines,
   e.g., when parentheses in expressions are set incorrectly.
   
\subsection{Validation of the type checker} \label{typer_checker_validation}
\begin{figure} 
\begin{center}
  \includegraphics[width=10.5cm]{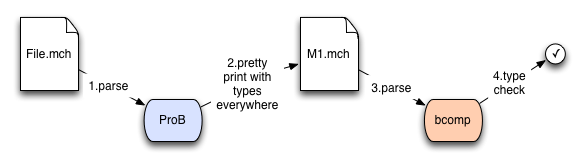} 
  \caption{Overview of the Type Checker Validation Process}\label{TypeCheckerValidation}
\end{center}
\end{figure}

  For the moment we also read in a large number of our regression test machines and pretty print the
   internal representation, this time with explicit typing information inserted.
  We now run this automatically generated file through the Atelier B parser and type checker
  (see Figure~\ref{TypeCheckerValidation}).
  With this we test whether the typing information inferred by our tool is compatible with
   the Atelier B type checker.
  (Of course, we cannot use this approach in cases where our type checker detects a type error.)
  Also, as the pretty printer only uses the minimal number of parentheses, we also ensure
   to some extent that our parser is compatible with the Atelier B parser (see below).
  Again, this validation can easily be applied to a large number of B machines.
  More importantly, it can be systematically applied
   to the particular B machines that \prob\ validates: provided the parser and pretty printer are correct,
    this gives us a guarantee that the typing information for those machines is correct.
  The latest version of \prob\ Tcl/Tk has a command to cross check the typing of the internal representation
    with Atelier B in this manner.

\paragraph*{Errors Detected}\label{frenchenglish:error}

During validation of the type checker we found several issues.
 (We did not detect errors in the type checker, but rather differences between our parser and the Atelier B parser {\tt bcomp}.)
For example, \prob\ accepts identifiers with single letter without warning
 and \prob\ accepts enumerated set elements that are used as identifiers while {\tt bcomp} does not.
We also detected that {\tt bcomp} reports a lexical error (``illegal token {\tt |-}'')
 if the vertical bar ({\tt |}) of a lambda abstraction is followed
 directly by the minus sign.
 \label{bcomp_err}
Most importantly, however, we found errors in the priority table of the english Atelier B ``B Language Reference Manual 1.8.6'' \cite{atelierb40}.
All in all there are 26 errors in the English reference manuals, upon which our pretty printer and BParser was based.
These issues have been fixed both in our parser and in the handbook.

\subsection{System Tests}

These tests exercise the entire \prob\ tool chain, including the parser, the type checker, the \prob\ interpreter and the \prob\ kernel.

\subsubsection{Regression Tests}
 \prob\ contains over 1100 regression tests which are made up of B models along with saved animation traces.
 These models are loaded, the saved animation traces are replayed, and the models are also run through the model checker.
 These tests have turned out to be extremely valuable in ensuring that a bug once fixed remains fixed.
 They are also very effective at uncovering errors in arbitrary parts of the system (e.g., the parser, type checker, the interpreter, the \prob\ kernel, etc.).
 The various features of \prob\ are also checked here: there are tests to check that \prob's visualization features work as expected,
   there are tests for \prob's test case generation algorithms, there are tests for the various constraint-based checks that \prob\ can perform, etc.

\subsubsection{Self-Model Check with Mathematical Laws}
  \label{it:model-check}%
 With this approach we use \prob's model checker to check itself, in particular the \prob\ kernel and the B interpreter.
 The idea is to formulate a wide variety of mathematical laws and then use the model checker to ensure that no counterexample to these laws can be found.
  
Concretely, \prob\ now checks itself for over 500 mathematical laws. In version 1.3.3 there were
 laws for booleans (39 laws), arithmetic laws (40 laws), laws for sets (81 laws), relations (189 laws), functions (73 laws) and sequences (61 laws),
 as well as some specific laws about integer ranges (24 laws) and the various basic integer sets (7 laws).
The number of laws is continuously being expanded.
Figure~\ref{SomeLaws} contains some of these laws about function.\footnote{Throughout the paper, we replaced B-style notation with mathematical notation for better comprehensibility.}

\begin{figure}
 \begin{scriptsize}
\begin{lstlisting}[mathescape]
   law1 ==  (dom(ff $\cup$ gg) = dom(ff) $\cup$ dom(gg));
   law2 ==  (ran(ff $\cup$ gg) = ran(ff) $\cup$ ran(gg));
   law3 ==  (dom(ff $\cap$ gg) $\subseteq$ dom(ff) $\cap$ dom(gg));
   law4 ==  (ran(ff $\cap$ gg) $\subseteq$ ran(ff) $\cap$ ran(gg));
   ....
 \end{lstlisting}
 \end{scriptsize}
  \caption{A small selection of the laws about B functions}\label{SomeLaws}
\end{figure}

This technique can be viewed as using the model checker to generate a very large number of tests from the mathematical laws.
The self-model check has been very effective at uncovering errors in the \prob\ kernel and interpreter.
Furthermore,  the self-model checking tests rely on every component of the \prob\
 execution environment working perfectly; otherwise
 a violation to a mathematical law could be found.
In addition to the \prob\ main code,
  the parser, type checker, Prolog compiler, hardware and operating system all have to work perfectly.
Indeed, we have identified a bug in our parser ({\tt FIN} was treated the same as {\tt FIN1}) using the self-model check.
Furthermore, we have even uncovered two
  bugs in the underlying SICStus Prolog compiler using self-model check:

\begin{itemize}
\item

The Prolog {\tt findall} did sometimes drop a list constructor, meaning that
 instead of {\tt [[]]} it sometimes returned {\tt []}.
In terms of B, this meant that instead of $\{\emptyset\}$ we received the empty set $\emptyset$.
This violated some of our mathematical laws about sets, notably the
 law \verb+POW1(SS) = POW(SS) - {{}}+.
This bug was reported to SICS and it was fixed in SICStus Prolog 4.0.2.
\item
A bug in the AVL library (notably in the predicate {\tt avl\_max} computing the maximum
 element of an AVL-tree) was found and reported to SICS.
 The bug was fixed in SICStus Prolog 4.0.5.
\end{itemize}

Note that these problems would not have been detected by validating or proving the code of
 \prob\ correct.
 It was essential to test the actual code of \prob\ along with the entire execution environment.
 The model checker together with the mathematical
  laws enabled this testing to be performed very effectively
  (see also Example~\ref{ex:double-evaluation} below).

We also applied some of these tests to other tools.
For example, we found various bugs in the first implementation of the JEB animator \cite{DBLP:conf/apsec/YangJS12}, 
 and the TLC \cite{DBLP:conf/charme/YuML99} model checker when applied to TLA translations of our B laws.


\section{Coverage}
\label{sec:coverage}

    The validation techniques explained above are complemented by code coverage analysis techniques.
In particular, we try to ensure that the unit tests (Section~\ref{it:unit})
 and the self-model checks (Section~\ref{it:model-check}) above cover
  all predicates and clauses of the \prob\ kernel.\footnote{We do not count the regression tests towards coverage, as there we basically compare
 against previous versions of \prob, which could in theory already contain a serious bug.}
We count a clause as covered if it has been executed successfully by at least one test case.
A predicate is regarded as covered if at least one of the clauses belonging to it is covered.
 
Initially, we have
 developed our own code coverage tool for SICStus Prolog.
The tool uses Prolog's term expansion facility to keep a record of which program points
 are covered.
As of version 4.2, SICStus Prolog now provides code coverage profiling, along with visualizations in
 Eclipse and Emacs.
The coverage is slightly less fine-grained than our original implementation, but it is much faster (as the profiling is directly
 supported by the Prolog abstract machine) and supports all of Prolog's features out of the box.
We have thus decided to switch to SICStus' code coverage profiler, but have implemented various backends to output
 the information in HTML, LaTeX, or other formats (see {\tt\small http://nightly.cobra.cs.uni-duesseldorf.de/coverage/}).
One of the backends features a colorized version of our source code, highlighting the coverage status.
With this backend, it is possible to inspect the source code itself, finding out exactly which parts of the
 code are covered.
This can be used to identify poorly tested source code and helps in adding missing test cases.
 
The code coverage statistics are also systematically computed by our continuous integration server (see Section~\ref{sec:jenkins}).
As of version 1.3.5 of \prob, we now also keep track of clauses which should not be covered as they relate
 to internal errors:
\begin{itemize}
  \item the {\tt\small error\_manager} module was extended to distinguish internal and normal errors.
   A normal error is, for example, a division by zero in a B machine. An internal error would be a typing error or
    unexpected failure within a piece of \prob\ Prolog code.
   Normal errors can be covered by tests; internal errors must not be coverable.
  \item a Prolog term expander keeps track of clauses and execution paths which relate to internal errors, and excludes those
   from the coverage statistics.
\end{itemize}


So far, the code coverage analysis has been very useful in extending our unit tests and mathematical laws.
As an example, 96.5 \% of the clauses of the {\tt kernel\_mappings} module are now covered
 by the unit tests and the mathematical laws. The only uncovered clauses relate to the
  Z compaction operator, and one error condition that was not triggered by the tests.
In summary, the code coverage has helped us write better tests and has allowed us to
  uncover a few undetected errors in the kernel.
  
However, even fully covered code might still contain errors.
Executing all parts of the source code does not necessarily mean that all paths were covered.
There might, for instance, be bugs that only occur if certain combinations of branches are executed.
Thus we can not rely solely on the coverage reports to judge the effectiveness of our test suite.
We experimented with artificially introduced bugs to evaluate how certain classes of programming errors can be found.
The following section will give a small overview.

 \subsection{Judging Effectiveness of Tests}
 
  In Section~\ref{testing} we already described the effectiveness of the current testing procedure
   at uncovering various errors in \prob\ and in the underlying compiler.
   We now check the effectiveness for artificially introduced errors.
   
   The tests in this subsection were conducted on \prob\ version 1.3.4-rc1, svn revision 9543.

%
 
   \begin{tabular}{|l|l|l|l|l|l|}
\hline
Error & Unit Tests & Regression  & Laws  \\
\hline
Replaced $\cap$ by set difference%
 & Passed & Passed & Failed\\
Subtle error in integer multiplication%
  & Failed & Passed & Failed%
 \\
Interchanged argument in not partial function test%
 & Passed & Passed  
 & Failed 
 \\
 Skip values when enumerating a variable
 & Failed & Passed  
 & Passed 
 \\
\hline
\end{tabular}

As you can see, all errors were detected and the mathematical laws checks detected most of the errors.
However, as checking the mathematical laws is based on searching for counter-examples, there is a certain class of errors that can not be found.
If due to a bug the search space is pruned, mathematical law checking will obviously not be able to find more counter-examples within the pruned area.
Therefore the mathematical law checks do not make the unit tests obsolete.

In the future, we plan to conduct a more extensive study about the effectiveness of our tests.


\section{Performance and Codespeed}
We run several benchmarks for \prob\ on a daily basis.
The results are stored using the codespeed web application\footnote{\url{https://github.com/tobami/codespeed/}} which allows us to compare selected revisions on selected benchmarks.
Furthermore, an unusual increase or decrease in performance is reported right on the front page.
Strictly speaking this does not add value to our test infrastructure, as the benchmarks do not report error or success.
However, in several occasions the existence of a bug in \prob\ lead to strange speed-ups or slowdowns, for instance, if a bug causes \prob\ to check every possible value of an existential quantification instead of stopping after finding a solution.
Therefore, monitoring the overall speed of our applications is a useful addition to testing.

In order to identify the particular revision causing unexpected benchmark results, we have found \verb|git bisect| to be useful.
It allows us to mark certain revisions as good (i.e. the last benchmarked revision that returned the expected results) and certain other revisions as bad (i.e. the revision showing strange behavior).
Git then performs an interactive binary search, checking out revisions in between until the one responsible has been found.

The process of comparing certain revisions of \prob\ has been partly automated by several scripts that control checkout, benchmarking and result reporting.


\section{Static Analysis}

Some bugs are hard to detect by testing but easy to detect with static analysis.
For instance, we had some calls to error handling predicates that were not used correctly, e.g., they were called with the wrong arity.
In most cases these calls happened in source code locations that are unreachable unless a previous bug occurs in the underlying execution engine. There were also bugs in the reachable parts of ProB that were not well tested, i.e., those parts that are of a more experimental nature. 

Prolog is not statically typed and therefore we cannot rely on a typechecker to identify incorrect calls to predicates, however it is very easy to implement an analysis for these sort of problems in Prolog by augmenting the interpreter with a term expander. 
Term expansion is very similar to macro expansion in Lisp.  
A term expander hooks into the loading process of a Prolog file and allows modification of a term while it is being loaded. 
In our case, we did not modify the terms, but extract information from it.
Hence, the expanded code is equal to the code used in production.
This allows us to relate all detected bugs directly to our code.
For instance, if we analyze the term \texttt{:- dynamic foo/1} in a module \texttt{bar}, we store the information that a predicate foo of arity 1 exists in the module bar and that it is a dynamic predicate. 
We also extract information about predicate and fact definitions, mode declarations, multifile and meta annotations, exported and imported predicates, blocking declarations for co-routines and most importantly calls. 
Strictly speaking this is not traditional static analysis, because Prolog directives such as \texttt{use\_module} are executed while loading the Prolog program. But the program is not run either, so we think it is appropriate to classify the approach as a static analysis.
To run the analyzer, we first load our term expander and then the entry point module of ProB. 
During loading the module, the term expander inspects all terms that are being loaded and stores the extracted information in a Prolog database.
The analyzer automatically detects calls where no definition is known at compile time. 
These calls are potential bugs. Using the tool, we were able to find 17 missing import statements, 9 cases where predicates are used with the wrong arity, 5 cases where the wrong predicate was called and 5 missing dynamic declarations. 
The missing imports did not actually cause wrong behavior at runtime because of a Sicstus Prolog implementation detail. As long as a module is loaded (even if it was loaded by any other module ), a predicate can be called using a fully qualified form \texttt{module:predicate(...)} at runtime. 
This is rather fragile because if for some reasons the module is not loaded the calls will break. Each module should declare its dependencies explicitly. 
The calls to predicates with wrong arity were also not critical because they were in dead code (if the Prolog engine works properly). 

Three cases where the wrong predicate was called were actual bugs in some experimental features of ProB. One feature was removed, the other two were fixed. The other two cases were dead code. The missing dynamic declarations are not dangerous in the current version of Sisctus Prolog, but there is no guarantee that this won't cause actual bugs in future versions. The problems we found using the tool are summarized in the following table:\\

\begin{tabular}{|l|c|c|c|}
\hline 
\rule[-1ex]{0pt}{2.5ex} {\bf Problem Type} & {\bf Fragile code/Bad practice} & {\bf Bug} & {\bf Dead code} \\ 
\hline 
\rule[-1ex]{0pt}{2.5ex} Import missing & 17 & 0 & 0 \\ 
\hline 
\rule[-1ex]{0pt}{2.5ex} Wrong arity & 0 & 0 & 9 \\ 
\hline 
\rule[-1ex]{0pt}{2.5ex} Wrong predicate & 0 & 3 & 2 \\ 
\hline 
\rule[-1ex]{0pt}{2.5ex} Missing dynamic delaration & 5 & 0 & 0 \\ 
\hline 
\end{tabular} \\

Because Prolog is a dynamic language, finding dead code is harder than in a static language such as Java. The compiler cannot know if a call is produced dynamically at runtime. This can easily lead to dead code, if we change our code. For instance, if we replace a call to predicate p by a call to predicate q, is p now dead code or is it used somewhere else? Using our analysis we get a list of predicates that are not being called in a static way. This clearly  produces false positives, but it is possible to manually inspect the candidates. We think that we can further improve the analysis by detecting typical dynamic usage pattern such as using a predicate in a higher order predicate, e.g. maplist.  

In addition to the static analysis, we are currently working on a tool that can be used to inspect the ProB source code. The tool should contain a code browser that allows to directly navigate beween caller and callee, search for predicates, filter according to specific criteria (e.g., parts of the name). This tool should also incorporate the documentation. We want to display certain properties of the source code (e.g., percentage of predicates that are documented or percentage of predicates with a mode declaration) to encourage the developers to increase the quality of the documentation and code.

Our custom tools are accompanied by Spider, an Eclipse based IDE specifically developed by SICStus for Prolog development. It features several static analyses, including calling mode and checks for potential non-determinacy and potential non-termination. In addition, there is an option to check for common mistakes and bad practice.


\section{Dynamic Runtime Analysis}

In this section we examine various ways we ensure consistency of \prob's output at runtime.

\subsection{Double chain}

One way to dramatically increase the confidence in \prob's output is to double check the result with a separately developed tool.
Thus far, this has been done with the following three tools:
\begin{enumerate}
 \item PredicateB. Together with the company ClearSy we have performed Data Validation tasks for Alstom \cite{DBLP:journals/corr/abs-1210-6815}.
   Here, the Java based tool PredicateB was used to double check \prob's output.
   Quite often, PredicateB took considerably more time than \prob\ to check the properties. Some properties needed to be rewritten to be suitable for PredicateB (whose constraint solving capabilities are much weaker).
   Overall, this approach was very successful and led to the development of the DTVT (Data Table Validation Tool) which integrates \prob\ and PredicateB.

 \item Ovado.
   In this work \cite{DBLP:journals/corr/abs-1210-7039}, the company Systerel is using the tool Ovado as primary tool, and uses \prob\ to double check Ovado's output.

 \item PyB.
   Finally, in recent work we are developing an independent tool written in Python to check B predicates.
   The idea is to feed \prob's output into PyB (in a more integrated fashion, e.g., passing witnesses for existential predicates) and to let PyB
    check the output.
   An independent companion paper about PyB has been submitted to this workshop.
\end{enumerate}

\subsection{Run Time Checking}
  The Prolog code contains a monitoring module which --- when turned on --- will 
  check pre- and post-conditions of certain predicate calls and also detect unexpected failures.
  Many kernel predicates also check for unexpected arguments.
  All of this overcomes the fact that Prolog has no static typing to some extent.
  So far we have not systematically relied on this feature; but we could turn the run time checking
   on during property validation.
 The only drawback is that the runtimes will be increased.
 

\subsection{Positive and Negative Evaluation} \label{double-evaluation}
   As already mentioned,  all properties and assertions were checked twice, both positively and negatively (see Figure~\ref{pos-neg-eval}).
   Indeed, \prob\ has two Prolog predicates to evaluate B predicates: one positive version which will succeed
    and enumerate solutions if the predicate is true and a negative version, which will succeed if the predicate is false
    and then enumerate solutions to the negation of the predicate.
    The reason for the existence of these two Prolog predicates is that Prolog's built-in negation is generally unsound and cannot be used to enumerate solutions in case of failure.
    For example, given the Prolog rule {\tt p(eq(X,Y)) :- X=Y.} the query {\tt not(p(eq(X,0)))} would fail, i.e.,
     one could erroneously conclude that there is no value {\tt X} which is different from {\tt 0}.
    In \prob, we have a dedicated predicate for the negation, which will
      suspend until it can determine the result correctly. For the above example, one could write
      {\tt not\_p(eq(X,Y)) :- dif(X,Y).}
  
 \begin{figure}
 \begin{center}
 \begin{tabular}{c|c|l}
 $P$ & $\neg P$ & Conclusion\\
 \hline
 {\tt true} & {\tt true} & bug in \prob\\
 {\tt true} & {\tt false} & $P$ is true\\
 {\tt false} & {\tt true} & $P$ is false\\
 {\tt false} & {\tt false} & $P$ is not well-defined\\
 {\tt true} & {\tt timeout} & $P$ is (probably) true\\
 {\tt false} & {\tt timeout} & $P$ is false or undefined\\
 {\tt timeout} & {\tt true} & $P$ is (probably) false\\
 {\tt timeout} & {\tt false} & $P$ is true or undefined\\
 {\tt timeout} & {\tt timeout} & $P$ is unknown\\
 \end{tabular}
 \caption{Positive and negative evaluation of predicates\label{pos-neg-eval}}
 \end{center}
 \end{figure}
 
\begin{figure}
  \begin{center}

    \includegraphics[width= 8cm]{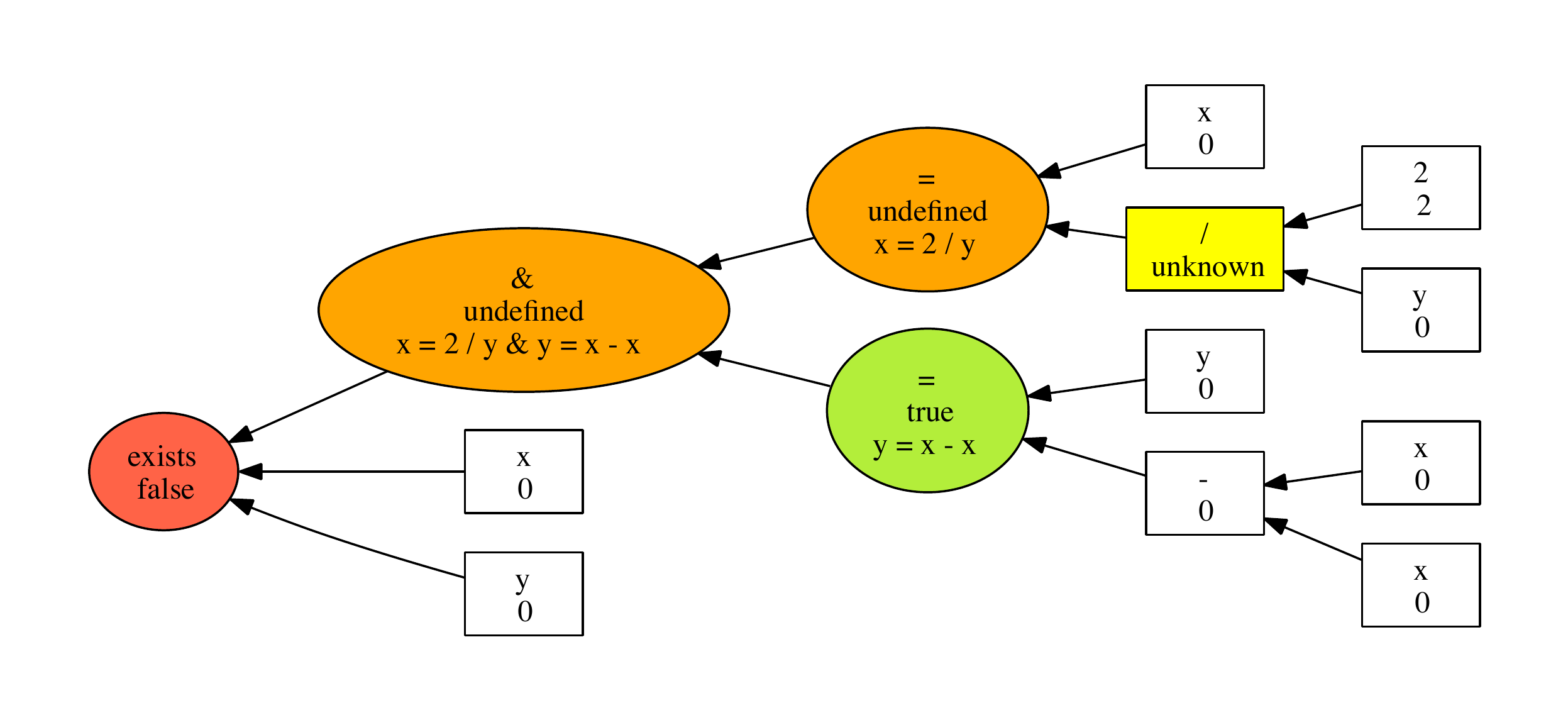}
  \end{center}
   \caption{Visualising an undefined property} 
\label{fig:undef}
\end{figure}

    With these two predicates we can uncover undefined predicates: 
  if for a given B predicate
     both the positive and negative Prolog predicates fail then the formula is undefined.
For example, the property {\tt x = 2/y \& y = x-x} over the constants {\tt x} and {\tt y} would be detected as being undefined,
 and would be visualized by our graphical formula viewer as in Figure~\ref{fig:undef} 
 (yellow and orange parts are undefined).
    
    In the context of validation, this approach has another advantage: for a formula to be classified as true the positive Prolog predicate must
     succeed {\em and\/} the negative Prolog predicate must fail, introducing a certain amount of redundancy (admittedly with common error modes).
     In fact, if both the positive and negative Prolog predicates would succeed for a particular B predicate then a bug in \prob\ would have been uncovered.
     If both fail, then either the B predicate is undefined or we have a bug in \prob.%
\footnote{Typically, \prob\ would also generate error messages for well-definedness problems.
Currently, there is however no guarantee that all well-definedness problems will result in an error message;
 the only certain way to detect them is to do both positive and negative evaluation.}
     
    This validation aspect can detect errors in the predicate evaluation parts of \prob,
      i.e., the treatment of the Boolean connectives
        $\vee$, $\wedge$, $\Rightarrow$, $\neg$, $\Leftrightarrow$,
        quantification $\forall$, $\exists$,
        and the
        various predicate operators such as $\in$, $\not\in$, =, $\neq$, $<$, ...
     This redundancy can not detect bugs inside expressions (e.g., $+$, $-$, ...)
       or substitutions (but the other validation aspects mentioned above can).
  
\begin{example} \label{ex:double-evaluation}
We want to illustrate the error detection of double evaluation by a simple example.
Take the following very simple B machine, with two assertions which should obviously be true.

\begin{footnotesize}
\begin{lstlisting}[mathescape]
      MACHINE DoubleEvaluationTest
      SETS ID={aa,bb}
      CONSTANTS iv 
      PROPERTIES iv $\in$ ID & iv $\neq$ bb
      ASSERTIONS iv $\in$ {aa}; iv $\notin$ {bb}
      END
\end{lstlisting}
\end{footnotesize}

Say that we introduce an error in the part of the source code that deals with testing membership of elements in a singleton set.
An element is a member of a singleton set if it is equal to the single element.
We replaced this test by its negation, and thus introduced
 an error into the \prob\ kernel.
When analyzing the assertions, we now get the following result, which tells us that
 there is a bug in the ProB kernel, because the assertion stating that {\tt iv} is not a member of the set {\tt \{bb\}} is both true and false at the same time:

\begin{footnotesize}
\begin{lstlisting}[mathescape]
      iv $\in$ {aa}
      == unknown
	
      iv $\notin$ {bb}
      == both_true_false
\end{lstlisting}
\end{footnotesize}

Also note that the self-model checking described in Section~\ref{it:model-check}
  immediately provides a counter example ({\tt SS=\{el1\}, TT=$\emptyset$})
  for several of the mathematical laws for sets, for example:

\begin{footnotesize}
\begin{lstlisting}[mathescape]
      {xx | xx $\in$ SS $\vee$ xx $\in$ TT } = SS $\cup$ TT
      == false
\end{lstlisting}
\end{footnotesize}
 
\end{example}


\section{Future Work and Discussion}

In the longer term we want to further increase the confidence in \prob's output so that it can eventually be used as a tool of class T3 within the railway norm EN 50128.

In addition to the development of the double chain PyB,
 there are several techniques that might help to increase the confidence in our primary validation tool even more.
One of these techniques is fuzzing or robustness testing.
The idea behind fuzzing would be to generate random input data and feed it to \prob.
As mentioned before, test coverage of error cases is hard to archive.
Random inputs of data could help us to identify cases in which some part of our tool chain crashes instead of reaching a controlled error state.

We could also further refine the results of the coverage analysis mentioned in Section \ref{sec:coverage}.
Instead of just observing which predicates were executed we could strive for coverage of possible execution paths or the evaluation of each branch.
However, this would again increase the complexity of the analysis.

Aside from improving the static checks strengthening the run time checks would be another reasonable approach.
With the increased usage of our code documentation tools more annotations will be added to the source code.
The correctness of several of these annotations, could be enforced at runtime.
One example, is the Prolog mode declaration which specifies which arguments are input arguments or output arguments respectively.



\section{Conclusion}

In summary, the following points increase our confidence in the proper working for \prob\ in general and for data validation in particular:
\begin{enumerate}
  \item repeated successful application on case studies
   (e.g., by Siemens on the lines of San Juan, Paris Line 1, Sao Paolo, Barcelona Line 9, ... \cite{DBLP:journals/fac/LeuschelFFP11}), discovering
    all known problems and in some  cases even discovering previously unknown problems in the data properties.
  \item extensive regression, unit-tests, automated unit-test generation,
    and validation via mathematical laws and model checking (see Section~\ref{testing}).
   These tests have even discovered several errors in the underlying Prolog compiler.
  \item coverage analysis of the testing, ensuring that all critical parts are covered by the tests;
    this has enabled us to detect a few more issues (and has prompted us to develop the automated unit-test generator).
  \item validation of the parser and type-checker via pretty-printing and cross-checking with the Atelier-B parser (bcomp) (see Section~\ref{parser_validation}).
   The internal representation of \prob\ for a particular model can be fed {\em at runtime} to bcomp, for additional validation that the type inference was properly conducted.
  \item double evaluation of predicates (see Section~\ref{double-evaluation}), to catch undefined predicates
   as well as errors in the \prob\ kernel and interpreter relating to predicates.
\end{enumerate}


\bibliographystyle{eptcs}

\bibliography{toolchain_validation}

\begin{thebibliography}{10}
\providecommand{\bibitemdeclare}[2]{}
\providecommand{\surnamestart}{}
\providecommand{\surnameend}{}
\providecommand{\urlprefix}{Available at }
\providecommand{\url}[1]{\texttt{#1}}
\providecommand{\href}[2]{\texttt{#2}}
\providecommand{\urlalt}[2]{\href{#1}{#2}}
\providecommand{\doi}[1]{doi:\urlalt{http://dx.doi.org/#1}{#1}}
\providecommand{\bibinfo}[2]{#2}

\bibitemdeclare{book}{Abrial:BBook}
\bibitem{Abrial:BBook}
\bibinfo{author}{Jean-Raymond \surnamestart Abrial\surnameend}
  (\bibinfo{year}{1996}): \emph{\bibinfo{title}{The {B}-Book}}.
\newblock \bibinfo{publisher}{Cambridge University Press},
  \doi{10.1017/CBO9780511624162}.

\bibitemdeclare{article}{rodinplatform}
\bibitem{rodinplatform}
\bibinfo{author}{Jean-Raymond \surnamestart Abrial\surnameend},
  \bibinfo{author}{Michael \surnamestart Butler\surnameend},
  \bibinfo{author}{Stefan \surnamestart Hallerstede\surnameend},
  \bibinfo{author}{Thai~Son \surnamestart Hoang\surnameend},
  \bibinfo{author}{Farhad \surnamestart Mehta\surnameend} \&
  \bibinfo{author}{Laurent \surnamestart Voisin\surnameend}
  (\bibinfo{year}{2010}): \emph{\bibinfo{title}{Rodin: an open toolset for
  modelling and reasoning in {Event-B}}}.
\newblock {\sl \bibinfo{journal}{STTT}}
  \bibinfo{volume}{12}(\bibinfo{number}{6}), pp. \bibinfo{pages}{447--466},
  \doi{10.1007/s10009-010-0145-y}.

\bibitemdeclare{article}{DBLP:journals/corr/abs-1210-7039}
\bibitem{DBLP:journals/corr/abs-1210-7039}
\bibinfo{author}{Frédéric \surnamestart Badeau\surnameend} \&
  \bibinfo{author}{Marielle \surnamestart Doche-Petit\surnameend}
  (\bibinfo{year}{2012}): \emph{\bibinfo{title}{Formal Data Validation with
  Event-B}}.
\newblock {\sl \bibinfo{journal}{CoRR}} \bibinfo{volume}{abs/1210.7039}.
\newblock
  \urlprefix\url{http://dblp.uni-trier.de/db/journals/corr/corr1210.html#abs-1210-7039}.

\bibitemdeclare{techreport}{EN50128}
\bibitem{EN50128}
\bibinfo{author}{\surnamestart CENELEC\surnameend} (\bibinfo{year}{2011}):
  \emph{\bibinfo{title}{Railway Applications -- Communication, signalling and
  processing systems -- Software for railway control and protection systems}}.
\newblock \bibinfo{type}{Technical Report} \bibinfo{number}{EN50128},
  \bibinfo{institution}{European Standard}.

\bibitemdeclare{manual}{atelierb40}
\bibitem{atelierb40}
\bibinfo{author}{\surnamestart ClearSy\surnameend} (\bibinfo{year}{2009}):
  \emph{\bibinfo{title}{Atelier {B}, User and Reference Manuals}}.
\newblock \bibinfo{address}{Aix-en-Provence, France}.
\newblock \bibinfo{note}{Available at {\tt http://www.atelierb.eu/}}.

\bibitemdeclare{mastersthesis}{gagnon:sablecc}
\bibitem{gagnon:sablecc}
\bibinfo{author}{Etienne \surnamestart Gagnon\surnameend}
  (\bibinfo{year}{1998}): \emph{\bibinfo{title}{{SableCC}, An Object-Oriented
  Compiler Framework}}.
\newblock Master's thesis, \bibinfo{school}{McGill University, Montreal,
  Canada}.
\newblock \bibinfo{note}{Available at {\tt http://www.sablecc.org}}.

\bibitemdeclare{article}{DBLP:journals/corr/abs-1210-6815}
\bibitem{DBLP:journals/corr/abs-1210-6815}
\bibinfo{author}{Thierry \surnamestart Lecomte\surnameend},
  \bibinfo{author}{Lilian \surnamestart Burdy\surnameend} \&
  \bibinfo{author}{Michael \surnamestart Leuschel\surnameend}
  (\bibinfo{year}{2012}): \emph{\bibinfo{title}{Formally Checking Large Data
  Sets in the Railways}}.
\newblock {\sl \bibinfo{journal}{CoRR}} \bibinfo{volume}{abs/1210.6815}.
\newblock
  \urlprefix\url{http://dblp.uni-trier.de/db/journals/corr/corr1210.html#abs-1210-6815}.

\bibitemdeclare{article}{DBLP:journals/fac/LeuschelFFP11}
\bibitem{DBLP:journals/fac/LeuschelFFP11}
\bibinfo{author}{Michael \surnamestart Leuschel\surnameend},
  \bibinfo{author}{J{\'e}r{\^o}me \surnamestart Falampin\surnameend},
  \bibinfo{author}{Fabian \surnamestart Fritz\surnameend} \&
  \bibinfo{author}{Daniel \surnamestart Plagge\surnameend}
  (\bibinfo{year}{2011}): \emph{\bibinfo{title}{Automated property verification
  for large scale {B} models with {ProB}}}.
\newblock {\sl \bibinfo{journal}{Formal Asp. Comput.}}
  \bibinfo{volume}{23}(\bibinfo{number}{6}), pp. \bibinfo{pages}{683--709},
  \doi{10.1007/s00165-010-0172-1}.

\bibitemdeclare{inproceedings}{DBLP:conf/apsec/YangJS12}
\bibitem{DBLP:conf/apsec/YangJS12}
\bibinfo{author}{Faqing \surnamestart Yang\surnameend},
  \bibinfo{author}{Jean-Pierre \surnamestart Jacquot\surnameend} \&
  \bibinfo{author}{Jeanine \surnamestart Souqui{\`e}res\surnameend}
  (\bibinfo{year}{2012}): \emph{\bibinfo{title}{The Case for Using Simulation
  to Validate Event-B Specifications}}.
\newblock In \bibinfo{editor}{Karl R. P.~H. \surnamestart Leung\surnameend} \&
  \bibinfo{editor}{Pornsiri \surnamestart Muenchaisri\surnameend}, editors:
  {\sl \bibinfo{booktitle}{APSEC}}, \bibinfo{publisher}{IEEE}, pp.
  \bibinfo{pages}{85--90}, \doi{10.1109/APSEC.2012.66}.

\bibitemdeclare{incollection}{DBLP:conf/charme/YuML99}
\bibitem{DBLP:conf/charme/YuML99}
\bibinfo{author}{Yuan \surnamestart Yu\surnameend}, \bibinfo{author}{Panagiotis
  \surnamestart Manolios\surnameend} \& \bibinfo{author}{Leslie \surnamestart
  Lamport\surnameend} (\bibinfo{year}{1999}): \emph{\bibinfo{title}{Model
  Checking TLA+ Specifications}}.
\newblock In \bibinfo{editor}{Laurence \surnamestart Pierre\surnameend} \&
  \bibinfo{editor}{Thomas \surnamestart Kropf\surnameend}, editors: {\sl
  \bibinfo{booktitle}{Correct Hardware Design and Verification Methods}}, {\sl
  \bibinfo{series}{Lecture Notes in Computer Science}} \bibinfo{volume}{1703},
  \bibinfo{publisher}{Springer Berlin Heidelberg}, pp. \bibinfo{pages}{54--66},
  \doi{10.1007/3-540-48153-2\_6}.

\end{thebibliography}

\end{document}